\documentclass{ws-ijgmmp}

\begin{document}

\markboth{}
{Thermodynamics of spacetime and unimodular gravity}

%
\catchline{}{}{}{}{}
%

\title{Thermodynamics of spacetime and unimodular gravity}

\author{A. Alonso-Serrano}

\address{Max-Planck-Institut  f\"ur  Gravitationsphysik  (Albert-Einstein-Institut)\\
Am M\"{u}hlenberg 1, 14476 Potsdam, Germany, \\
email: ana.alonso.serrano@aei.mpg.de}

\author{M. Li\v{s}ka}

\address{Institute of Theoretical Physics, Faculty of Mathematics and Physics, Charles University\\
V Hole\v{s}ovi\v{c}k\'{a}ch 2, 180 00 Prague 8, Czech Republic}
\address{Max-Planck-Institut  f\"ur  Gravitationsphysik  (Albert-Einstein-Institut)\\
Am M\"{u}hlenberg 1, Potsdam, Germany, \\
email: liska.mk@seznam.cz}

\maketitle


\begin{abstract}
In this review we discuss emergence of unimodular gravity (or, more precisely, Weyl transverse gravity) from thermodynamics of spacetime. By analyzing three different ways to obtain gravitational equations of motion by thermodynamic arguments, we show that the results point to unimodular rather than fully diffeomorphism invariant theories and that this is true even for modified gravity. The unimodular character of dynamics is especially evident from the status of cosmological constant and energy-momentum conservation.
\end{abstract}


\section{Introduction}	

The relation between thermodynamics and gravitational physics was first described for the case of black holes. It was found that the Einstein equations imply a set of laws of black hole dynamics analogous to those of thermodynamics~\cite{Bardeen:1973} and that black holes posses entropy proportional to the horizon area~\cite{Bekenstein:1973}. Furthermore, due to quantum effects they6 emit black body radiation and, therefore, posses a finite temperature~\cite{Hawking:1975}. Analogous thermodynamic properties have been later described for other geometries with causal horizons, e.g. de Sitter spacetimes~\cite{Gibbons:1977}, Rindler wedges~\cite{Jacobson:2003} and causal diamonds~\cite{Jacobson:2019}.

A novel perspective on the connection between thermodynamics and gravity was offered by the idea that equations governing gravitational dynamics can be recovered from thermodynamic arguments~\cite{Jacobson:1995}. To do so, one needs to construct a local (approximate) causal horizon in every spacetime point and assign to them temperature and entropy. This approach has been later refined~\cite{Chirco:2010,Baccetti:2013} and even shown to work for some alternative theories of gravity~\cite{Eling:2006,Padmanabhan:2010,Jacobson:2012,Bueno:2017,Svesko:2018,Svesko:2019} and for the cases when gravity is sourced by quantum fields~\cite{Bueno:2017,Svesko:2019,Jacobson:2015}.

The aim of this review is to argue that thermodynamics of spacetime leads naturally to unimodular theories of gravity rather than fully diffeomorphism invariant ones. Unimodular gravity (UG) is classically equivalent to general relativity (GR), but its symmetry group is reduced to transverse (metric determinant preserving) diffeomorphisms. Similarly, there exists a unimodular version of any local, diffeomorphism invariant theory of gravity. One of the appeals of UG lies in the fact, that vacuum energy does not couple to gravity, solving some (but not all) of the problems related to the value of the cosmological constant~\cite{Fiol:2010,Bufalo:2015,Padilla:2015,Barcelo:2018}. Instead of a fully fledged theory of gravity, UG can also be understood as gauge fixed Weyl transverse gravity (WTG), a theory invariant under transverse diffeomorphisms and Weyl transformations. Just like UG, WTG is classically equivalent to general relativity (GR).

Various aspects of the relation of thermodynamics of spacetime and UG have been noted in several works~\cite{Padmanabhan:2010,Tiwari:2006,Alonso:2020a,Alonso:2020b}. Here, we try to provide a broad perspective on this connection by discussing the emergence of UG in three distinct thermodynamic derivations. The most straightforward one is based on the seminal Jacobson's paper concerning thermodynamics of Rindler wedges~\cite{Jacobson:1995} (with some later improvements taken into account~\cite{Chirco:2010,Baccetti:2013}). The second one obtains gravitational dynamics from thermodynamics of local causal diamonds. As it turns out, the properties of these objects allow us to derive equations of motion for a wide class of modified theories of gravity~\cite{Svesko:2018,Svesko:2019}. The last approach also considers causal diamonds, but describes the sources of gravity as quantum fields and, thus, leads to semiclassical gravitational dynamic~\cite{Jacobson:2015}. In all the cases, we find distinctly unimodular behavior of the resulting gravitational dynamics.

The paper is organized as follows. In section~\ref{UG} we briefly introduce UG and WTG. Section~\ref{TD} concerns emergence of UG (or WTG) from thermodynamics of spacetime. Finally, in section~\ref{discussion} we sum up the results and discuss possible future developments.

For simplicity of notation we consider four spacetime dimensions although a generalization to an arbitrary dimension is trivial. We work with metric signature $(-,+,+,+)$. Definitions of the curvature-related quantities follow~\cite{MTW}. Lower case Greek letters denote abstract spacetime indices. Unless otherwise explicitly stated, we use the SI units.

\section{Unimodular and Weyl transverse gravity}
\label{UG}

The origin of UG can be traced all the way back to Einstein~\cite{Einstein:1919}. The basic idea is to restrict the full diffeomorphism invariance of GR only to transverse diffeomorphisms, i.e., those generated by an arbitrary divergence-free vector field $\xi^{\mu}$,
\begin{align}
\nonumber g'_{\mu\nu}&=g_{\mu\nu}+2\xi_{(\mu;\nu)},  \\
\xi^{\mu}_{\;\:;\mu}&=0.
\end{align}
Such diffeomorphisms do not change the metric determinant, which one keeps fixed to $\sqrt{-g}=\omega_0$, where $\omega_0$ is an arbitrary real number (Einstein's original choice was $g=-1$). While Einstein proposed this restriction on the level of equations of motion (EoMs), it can be implemented in variational formulation of the gravitational dynamics. The simplest action for UG reads~\cite{Bufalo:2015}
\begin{equation}
\label{action}
S=\int_{\Omega}\left[\frac{c^4}{16\pi G}\left(R-2\bar{\Lambda}\right)+\mathcal{L}_{\text{matter}}\right]\omega_0\text{d}^4x,
\end{equation}
where $\Omega$ is a spacetime manifold, $\bar{\Lambda}$ is a constant and $L_{\text{matter}}$ denotes the matter Lagrangian. When varying the action with respect to the metric, the allowed variations are only those with $\delta g=0$. Otherwise, they would break the fixed determinant condition, $\sqrt{-g}=\omega_0$. Varying the UG action with respect to the metric under this condition yields traceless EoMs
\begin{equation}
R_{\mu\nu}-\frac{1}{4}Rg_{\mu\nu}=\frac{8\pi G}{c^4}\left(T_{\mu\nu}-\frac{1}{4}Tg_{\mu\nu}\right).
\end{equation}
Clearly, constant $\bar{\Lambda}$ plays no role whatsoever in gravitational dynamics. Hence, we simply set $\bar{\Lambda}=0$ in the following.

There are several ways to refine this basic treatment of UG. First, we may introduce a more general position-dependent background (nondynamical) spacetime volume element, $\omega\left(x\right)\text{d}^4x$, where $\omega$ is a strictly positive function. Second, rather than being fixed ``by hand'', the unimodular condition, $\sqrt{-g}=\omega$, can be implemented in the action via a Lagrange multiplier, $\lambda$. However, the action then acquires an additional ambiguity~\cite{Finkelstein:2001}
\begin{equation}
\label{action 2}
S=\int_{\Omega}\left[\frac{c^4}{16\pi G}R+\frac{c^4}{16\pi G}\lambda\left(\frac{\omega}{\sqrt{-g}}-1\right)+\mathcal{L}_{\text{matter}}+\left(\frac{\omega}{\sqrt{-g}}-1\right)l_{\text{M}}\right]\sqrt{-g}\text{d}^4x
\end{equation}
where $l_{\text{M}}$ is some undetermined function of the metric and matter variables. This term appears because we have some freedom in extending the action from the unimodular gauge to metrics with arbitrary determinants. To fully describe the gravitational dynamics we must vary the action with respect to both the metric and the Lagrange multiplier. Variation with respect to $\lambda$ yields
\begin{equation}
\sqrt{-g}=\omega,
\end{equation}
i.e., it enforces the unimodular condition on the metric. To find gravitational EoMs, we now vary the action with respect to metric without any restrictions and find 
\begin{equation}
R_{\mu\nu}-\frac{1}{2}Rg_{\mu\nu}+\frac{1}{2}\lambda g_{\mu\nu}=\frac{8\pi G}{c^4}\left(T_{\mu\nu}-\frac{1}{2}l_{\text{M}}g_{\mu\nu}\right).
\end{equation}
We can see that the energy-momentum tensor is not uniquely defined, as it can be shifted by a term of the form $l_{\text{M}}g_{\mu\nu}/2$, with $l_{\text{M}}$ being an arbitrary scalar function. Likewise, the Lagrange multiplier $\lambda$ on the left hand side is completely undetermined by the gravitational dynamics. To eliminate both ambiguous terms, one may express $\lambda$ from the trace of EoMs
\begin{equation}
\lambda=\frac{1}{2}R+\frac{8\pi G}{c^4}\left(\frac{1}{2}T-l_{\text{M}}\right).
\end{equation}
Substituting this result into EoMs yields
\begin{equation}
R_{\mu\nu}-\frac{1}{4}Rg_{\mu\nu}=\frac{8\pi G}{c^4}\left(T_{\mu\nu}-\frac{1}{4}Tg_{\mu\nu}\right).
\end{equation}
Hence, the form of UG action with a Lagrangian multiplier~\eqref{action 2} unambiguously determines only the traceless part of EoMs, just like the simpler action~\eqref{action}.

Let us now discuss the question of local energy-momentum conservation. In GR, diffeomorphism invariance of the matter Lagrangian implies $T_{\mu\;\:;\nu}^{\;\:\nu}=0$ and, therefore, energy-momentum is locally conserved. In contrast, transverse diffeomorphism invariance of UG does not suffice to ensure divergence-free energy momentum tensor. The question then is which mechanisms, if any, can lead to energy-momentum non-conservation in UG. First option that comes to mind is that divergence of the energy-momentum tensor might be proportional to the gradient of either the Lagrange multiplier, $\lambda$, or the ambiguity in UG action, $l_{\text{M}}$.  However, as we have already noted, neither $\lambda$ nor $l_{\text{M}}$ affect the traceless part of the energy-momentum tensor. Furthermore, they do not enter the EoMs for matter fields (variations of $l_{\text{M}}$ with respect to matter fields are all multiplied by $\omega-\sqrt{-g}$ and, thus, vanish in the unimodular gauge). To sum up, $\lambda$ and $l_{\text{M}}$ influence neither gravitational nor matter EoMs in any way and they have no physical effect whatsoever, at least on the level of classical dynamics~\cite{Finkelstein:2001}. Thus, we can just disregard them in the following.

Another possibility is that the matter Lagrangian, $\mathcal{L}_{\text{matter}}$, may be such that energy-momentum is not locally conserved (this may happen, e.g. in non-unitary quantum dynamics~\cite{Josset:2017} or due to diffusion of energy from matter to spacetime degrees of freedom~\cite{Perez:2018}). Nevertheless, invariance with respect to transverse diffeomorphisms constrains divergence of the energy-momentum tensor to be~\cite{Alvarez:2013}
\begin{equation}
\frac{8\pi G}{c^4}T_{\mu\;\:;\nu}^{\;\:\nu}=\mathcal{J}_{;\mu},	
\end{equation}
for some scalar function $\mathcal{J}$. This function measures the local non-conservation of energy-momentum which is satisfied only if $T_{\mu\;\:;\nu}^{\;\:\nu}=0$, i.e., $\mathcal{J}=0$. However, one can also describe the gravitational dynamics in terms of a new energy-momentum tensor
\begin{equation}
T'_{\mu\nu}=T_{\mu\nu}-\frac{8\pi G}{c^4}\mathcal{J}g_{\mu\nu},
\end{equation}
that satisfies $T_{\mu\;\:;\nu}^{'\;\nu}=0$. Notably, the traceless parts of $T_{\mu\nu}$ and $T'_{\mu\nu}$ are equal
\begin{equation}
T'_{\mu\nu}-\frac{1}{4}T'g_{\mu\nu}=T_{\mu\nu}-\frac{c^4}{8\pi G}\mathcal{J}g_{\mu\nu}-\frac{1}{4}Tg_{\mu\nu}+\frac{c^4}{8\pi G}\mathcal{J}g_{\mu\nu}=T_{\mu\nu}-\frac{1}{4}Tg_{\mu\nu}.
\end{equation}
Since varying the UG action unambiguously determines only the traceless part of the energy-momentum tensor, $T_{\mu\nu}$ and $T'_{\mu\nu}$ can be used interchangeably on the level of gravitational EoMs. Regardless of which energy-momentum tensor one chooses, taking a covariant divergence of the traceless EoMs and invoking the contracted Bianchi identities yields
\begin{equation}
\nabla_{\mu}\left(\mathcal{J}+\frac{1}{4}R-\frac{1}{4}\frac{8\pi G}{c^4}T\right)=0,
\end{equation}
and, consequently,
\begin{equation}
\mathcal{J}+\frac{1}{4}R-\frac{1}{4}\frac{8\pi G}{c^4}T=\Lambda,
\end{equation}
where $\Lambda$ is an arbitrary integration constant. Substituting this result back into the traceless EoMs leads to
\begin{equation}
\label{Einstein}
R_{\mu\nu}-\frac{1}{2}Rg_{\mu\nu}+\Lambda g_{\mu\nu}=\frac{8\pi G}{c^4}T_{\mu\nu}-\mathcal{J}g_{\mu\nu}.
\end{equation}
These equations are of the same form as Einstein equations with a divergence-free energy-momentum tensor $T'_{\mu\nu}=T_{\mu\nu}-\mathcal{J}g_{\mu\nu}$. However, the metric is still restricted to be unimodular. Moreover, the role of the cosmological constant is played by an integration constant, $\Lambda$. In contrast with GR, $\Lambda$ is not related to any fixed constant parameter in the Lagrangian. Instead, the value of $\Lambda$ is arbitrary and free to vary between solutions.

Now consider UG sourced by an energy-momentum tensor, $T_{\mu\nu}$, derived from some matter Lagrangian, $L_{\text{matter}}$, which is not divergence-free, i.e, $\left(8\pi G/c^4\right)T_{\mu\;\:;\nu}^{\;\:\nu}=\mathcal{J}_{;\mu}\ne0$. It has been argued that $\mathcal{J}$ can then be understood as a non-constant contribution to dark energy (appearing in addition to the constant one coming from the integration constant $\Lambda$), offering a possible explanation for accelerating expansion of the universe within the context of UG~\cite{Josset:2017,Perez:2018,Tiwari:1993,Tiwari:2003}. However, as we have seen, divergence-free energy-momentum tensor $T'_{\mu\nu}=T_{\mu\nu}-\left(c^4/8\pi G\right)\mathcal{J}g_{\mu\nu}$ leads to the same gravitational dynamics. Since $T'_{\mu\nu}$ represents a legitimate energy-momentum tensor both in UG and GR, it might appear, on the one hand, that UG provides no mechanism for non-constant dark energy that is not already possible in GR. On the other hand, energy-momentum tensor $T'_{\mu\nu}$ corresponds to some matter Lagrangian, $L'_{\text{matter}}$, different from $L_{\text{matter}}$. Variations of $L'_{\text{matter}}$ and $L_{\text{matter}}$ with respect to matter variables lead to different EoMs for matter fields and, therefore, are in principle physically distinguishable. Whereas UG allows matter sources described by either $L'_{\text{matter}}$ or $L_{\text{matter}}$, only $L'_{\text{matter}}$ can be coupled to gravity in GR. Hence, UG might indeed provide a novel explanation for dark energy, as suggested in the literature.

To conclude, let us stress that, besides the status of the cosmological constant and local energy-momentum conservation, UG is classically fully equivalent to GR. However, their equivalence on the quantum level has been studied in a number of works, with differing conclusions~\cite{Fiol:2010,Bufalo:2015,Padilla:2015,Kugo:2021} and remains an open question.

There are several ways to extend the basic UG action~\eqref{action}. On one side, the full diffeomorphism invariance may be recovered by introducing new degrees of freedom besides the metric~\cite{Henneaux:1989}. However, since such degrees of freedom cannot (to our best knowledge) be recovered from thermodynamics of spacetime, we do not discuss approaches of this kind further. On the other side, one can understand UG as a theory of gravity invariant under both transverse diffeomorphisms and Weyl transformations~\cite{Barcelo:2018,Carballo:2015,Alvarez:2006} restricted to the unimodular gauge. To see this, we introduce an auxiliary metric
\begin{equation}
\tilde{g}_{\mu\nu}=\left(\frac{\omega}{\sqrt{-g}}\right)^{\frac{1}{2}}g_{\mu\nu},
\end{equation}
so that $\sqrt{-\tilde{g}}=\omega$ and $\tilde{g}_{\mu\nu}$ is by construction a unimodular metric. Note that the transformation from $g_{\mu\nu}$ to $\tilde{g}_{\mu\nu}$ is clearly non-invertible (we lose one degree of freedom corresponding to the determinant). In terms of the auxiliary metric, the UG action reads
\begin{equation}
\label{action 3}
S=\int_{\Omega}\left(\frac{c^4}{16\pi G}\tilde{R}+\mathcal{L}_{\text{matter}}\right)\omega\text{d}^4x,
\end{equation}
where $\tilde{R}$ is an auxiliary Ricci scalar constructed from $\tilde{g}_{\mu\nu}$ and the corresponding Levi-Civita connection (i.e., a torsion-free connection obeying $\tilde{\nabla}_{\rho}\tilde{g}_{\mu\nu}=0$). The action written in terms of $\tilde{g}_{\mu\nu}$ is clearly invariant under Weyl transformations
\begin{equation}
g'_{\mu\nu}=e^{2\sigma}g_{\mu\nu},
\end{equation}
where $\sigma$ is some scalar function, as well as under transverse diffeomorphisms (defined with respect to the auxiliary Levi-Civita connection)
\begin{align}
\nonumber g'_{\mu\nu}&=g_{\mu\nu}+2\xi_{(\mu;\nu)},  \\
\tilde{\nabla}_{\mu}\xi^{\mu}&=0,
\end{align}
but they are not fully diffeomorphism invariant. By introducing the Weyl invariance, we expanded UG into a theory known as WTG~\cite{Barcelo:2018,Carballo:2015,Alvarez:2006}. From this perspective, UG represents a gauge-fixed form of WTG, selected by the condition $\sqrt{-g}=\omega$ (or, equivalently, by treating the auxiliary metric $\tilde{g}_{\mu\nu}$ as the fundamental dynamical variable).

To obtain the EoMs, vary action~\eqref{action 3} with respect to the dynamical metric, $g^{\mu\nu}$. The result are traceless, WTDiff invariant equations
\begin{equation}
\tilde{R}_{\mu\nu}-\frac{1}{4}\tilde{R}\tilde{g}_{\mu\nu}=\frac{8\pi G}{c^4}\left(T_{\mu\nu}-\frac{1}{4}Tg_{\mu\nu}\right).
\end{equation}
Discussion of local energy-momentum conservation proceeds along the same lines as for UG with two notable differences. First, since WTG is not restricted to the unimodular gauge, neither the Lagrange multiplier $\lambda$ nor the action ambiguity $l_{\text{M}}$ appear. Second, one must be careful to always work with Weyl invariant expressions and, in particular, use the Weyl covariant derivative, $\tilde{\nabla}_{\mu}$. The energy-momentum tensor obeys
\begin{equation}
\frac{8\pi G}{c^4}\tilde{\nabla}_{\nu}\left[\left(\frac{\sqrt{-g}}{\omega}\right)^{\frac{2}{n}}T_{\mu}^{\;\:\nu}\right]=\tilde{\nabla}_{\mu}\mathcal{J},
\end{equation}
for some function $\mathcal{J}$ measuring the local energy-momentum non-conservation. The contracted Bianchi identities then again allow us to rewrite the EoMs in an Einstein-like form
\begin{equation}
\tilde{R}_{\mu\nu}-\frac{1}{2}\tilde{R}\tilde{g}_{\mu\nu}+\Lambda\tilde{g}_{\mu\nu}=\frac{8\pi G}{c^4}T_{\mu\nu}-\mathcal{J}\tilde{g}_{\mu\nu}.
\end{equation}
Just like in UG, the cosmological constant $\Lambda$ appears as an integration constant and $\mathcal{J}$ can be interpreted as a non-constant contribution to dark energy (with the same caveats as in UG).

One of the main reasons for interest in UG and WTG lies in their potential to solve some of the issues associated with the cosmological constant in GR. To close our review of these theories, we provide a very brief discussion of this topic. More detailed treatment can be found, e.g. in~\cite{Weinberg:1989,Polchinski:2006,Burgess:2013}.

The advantage of UG and WTG over GR lies in different interplay between vacuum energy and gravitational fields. Quantum vacuum is populated by virtual particles represented in terms of Feynman diagrams by vacuum bubbles. These diagrams have no physical effect in flat spacetime quantum field theory, but this changes when gravity is taken into account. In GR, the dependence of spacetime volume on the metric leads to coupling of vacuum bubbles to gravity. Then, there appears gravitating vacuum energy-momentum tensor of the form $Cg_{\mu\nu}$, for some constant $C$, which behaves like a cosmological constant. This contribution is many orders of magnitude higher than the experimental value of the cosmological constant\footnote{Notably, this vacuum energy contribution to cosmological constant is ruled out not only by cosmological models, but even by solar system experiments~\cite{Kagramanova:2006}}~\cite{Barcelo:2018}. More worryingly, even if one fixes the cosmological constant to the required value on the tree level, every higher-order loop correction we add will again lead to contributions much larger than the experimentally found cosmological constant and will require further fine-tuning. In other words, the value of the cosmological constant is radiatively unstable in GR~\cite{Carballo:2015}. This challenges the rationale behind the effective field theory approach used in such calculations, i.e., that high energy physics does not significantly alter results of low energy experiments.

The situation is very different in UG and WTG. Since both theories introduce a metric-independent spacetime volume element, the vacuum energy does not couple to gravity~\cite{Barcelo:2018}. This is most evident from the fact that the EoMs are invariant under a simultaneous shift of the energy-momentum tensor by $Cg_{\mu\nu}$ and the cosmological constant by $C$, where $C$ can be any constant.\footnote{This symmetry does not occur in GR, as $\Lambda$ is a fixed parameter in the Lagrangian whose value cannot be shifted. It is present in UG and WTG because $\Lambda$ appears as an arbitrary integration constant.} Since $Cg_{\mu\nu}$ is precisely the form of vacuum energy-momentum tensor, its contribution is irrelevant in UG and WTG.

WTG offers an even stronger result. Radiative instability of the cosmological constant generically occur if the gravitational Lagrangian contains any term of the form $\Lambda\sqrt{-g}$, with $\Lambda$ being a constant~\cite{Barcelo:2018}. Presence of such terms is forbidden in the classical WTG due to its Weyl symmetry. Moreover, there are no quantum anomalies associated with this symmetry~\cite{Carballo:2015,Alvarez:2013b} and, thus, these terms remain forbidden even on the quantum level (this behavior crucially depends on breaking the invariance with respect to longitudinal diffeomorphisms~\cite{Barcelo:2018,Carballo:2015}). In conclusion, the value of cosmological constant is radiatively stable in WTG.

\section{Gravitational dynamics from thermodynamics}
\label{TD}

In the following, we discuss three distinct ways to obtain gravitational dynamics from thermodynamic arguments. The first one is similar to the original derivation presented by Jacobson for local Rindler wedges. The second works with causal diamonds and has the advantage of being applicable even to modified theories of gravity. The last one also uses causal diamonds, but treats the sources of gravity as quantum fields, leading to semiclassical gravitational dynamics. As we will see, in every case we recover a unimodular theory of gravity rather than a fully diffeomorphism invariant one.

\subsection{Thermodynamics of local Rindler wedges}
\label{Jacobson}

\begin{figure}[tbp]
\centering
\includegraphics[width=.3\textwidth,origin=c,trim={11.5cm 4.6cm 12.2cm 2.0cm},clip]{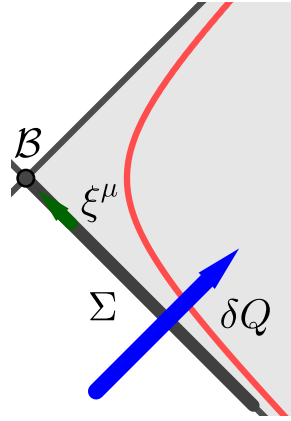}
\caption{\label{Rindler} A sketch of the local approximate Rindler wedge. The bifurcate spacelike 2-surface $\mathcal{B}$ is denoted by a small black circle, red line is an example trajectory of a uniformly accelerated observer perceiving the horizon. The selected part of one branch of the past causal horizon $\Sigma$ is represented by a thick black line. The blue arrow shows the direction of the heat flux across the horizon and the green one represents null vector $\xi^{\mu}$ normal to $\Sigma$.}
\end{figure}

First, we review a version of the original Jacobson's derivation~\cite{Jacobson:1995}. The crucial ingredient of any thermodynamic derivation is a local approximate causal horizon that can be constructed in every regular spacetime point. Here, we consider the simplest option, a local approximate Rindler wedge. In an arbitrary spacetime point $P$ introduce a locally flat coordinate system. Choose a small (with respect to the local curvature length scale) patch $\mathcal{B}$ of a spacelike 2-surface passing through $P$. Then select a small piece of the one branch of the null boundary of the causal past of $\mathcal{B}$ and denote it by $\Sigma$. The construction is sketched in Fig.~\ref{Rindler}.

Now, consider a timelike hypersurface $\Sigma'$ swept out by wordlines of a class of observers with constant acceleration $a$, so that $\Sigma'$ approaches $\Sigma$ in the limit of \mbox{$a\to\infty$}. If the length of wordlines forming $\Sigma'$ is much larger than $c^2/a$, this hypersurface has a well-defined Unruh temperature, $T_{\text{U}}=\hbar a/2\pi k_Bc$~\cite{Barbado:2012}. The validity of the Unruh temperature formula further requires that the local inertial vacuum state of quantum fields is well approximated by the standard Minkowski vacuum. Assuming this is equivalent to the validity of Einstein equivalence principle\footnote{For the statements and hierarchy of various formulations of the equivalence principle, see~\cite{Casola:2015}.}.

In the presence of matter, one can state the heat flux $\delta Q$ across $\Sigma$ in terms of the energy-momentum tensor. Then, invoking the equilibrium Clausius relation, $\text{d}S_{\text{Clausius}}=\delta Q/T_{\text{U}}$, one obtains an expression for the Clausius entropy flux across $\Sigma'$. Taking the limit of $a\to\infty$ then yields the Clausius entropy flux across the null hypersurface $\Sigma$
\begin{equation}
\Delta S_{\text{Clausius}}=\frac{2\pi k_B}{\hbar c}\int_{\Sigma}\lambda T_{\mu\nu}\xi^{\mu}\xi^{\nu}\text{d}^3\Sigma,
\end{equation}
where $\lambda$ denotes a parameter along the (approximate) geodesics forming $\Sigma$ (different parametrization of each of the geodesics is allowed) and $\xi^{\mu}$ is a future-oriented null vector field orthogonal to $\Sigma$. This formula is not restricted to Rindler wedges but holds for essentially any bifurcate null surfaces for which the spacetime curvature effects can be neglected~\cite{Baccetti:2013}. Notably, since its definition relies on the Unruh effect, Clausius entropy is semiclassical rather than classical, which is evident from the presence of $\hbar$ in the equation.

In addition to the Clausius entropy associated with the matter flux, we assume that the local Rindler horizon possesses entropy proportional to the area of its cross-section, $S_{\text{horizon}}=\eta\mathcal{A}\left(\mathcal{B}\right)$, where $\eta$ is constant throughout the spacetime (this assumption implicitly demands that the strong equivalence principle holds~\cite{Chirco:2010,Alonso:2020a}). A natural interpretation of this entropy in terms of quantum entanglement has been presented in the context of the search for microscopic origin of black hole entropy~\cite{Sorkin:1986,Srednicki:1993,Solodukhin:2011}. The idea is that an observer on one side of any causal horizon measures entanglement entropy proportional to its area. This entanglement is often assumed to be the only source of horizon's entropy in the context of thermodynamics of spacetime~\cite{Jacobson:1995,Chirco:2010}. Nevertheless, we stress that our reasoning does not depend on the way we interpret the horizon's entropy.

Therefore, finding an expression for $\Delta S_{\text{horizon}}$ is simply a matter of describing evolution of the horizon's cross-section area. Its change can be described in terms of expansion $\theta$ of the geodesic congruence forming $\Sigma$, yielding for the change of entropy
\begin{equation}
\Delta S_{\text{horizon}}=\eta\int_{\Sigma}\frac{\text{d}\theta}{\text{d}\lambda}\text{d}^3\Sigma.
\end{equation}
To evaluate $\text{d}\theta/\text{d}\lambda$ we use the Raychaudhuri equation (to simplify the discussion, we assume vanishing expansion and shear at $\mathcal{B}$, but this assumption can be relaxed~\cite{Chirco:2010}). Let us stress that this equation is a geometric identity completely independent of gravitational dynamics and we are thus free to invoke it without making a circular argument. The resulting expression for the entropy change reads
\begin{equation}
\Delta S_{\text{horizon}}=-\eta\int_{\Sigma}\lambda R_{\mu\nu}\xi^{\mu}\xi^{\nu}\text{d}^3\Sigma.
\end{equation}

If we assume that the local Rindler wedge is in thermal equilibrium, the Clausius entropy crossing the horizon must be compensated by the change of its (entanglement) entropy
\begin{equation}
0=\Delta S_{\text{Clausius}}+\Delta S_{\text{horizon}}=\int_{\Sigma}\lambda\left(\frac{2\pi k_B}{\hbar c}T_{\mu\nu}-\eta R_{\mu\nu}\right)\xi^{\mu}\xi^{\nu}\text{d}^3\Sigma.
\end{equation}
Since this identity in spacetime point $P$ holds for any null vector $\xi^{\mu}$ (one just needs to consider all the possible Rindler wedges crossing $P$), the bracket inside the integral needs to be equal to a term of the form $\eta\Phi g_{\mu\nu}$, where $\Phi$ is some scalar function (such a term can appear because $g_{\mu\nu}\xi^{\mu}\xi^{\nu}=0$). The strong equivalence principle guarantees that we obtain the same condition for every spacetime point $P$. Hence, throughout the spacetime it holds
\begin{equation}
R_{\mu\nu}+\Phi g_{\mu\nu}=\frac{2\pi k_B}{\eta\hbar c}T_{\mu\nu}.
\end{equation}
At this point, most of the treatments in the literature impose the local energy-momentum conservation, i.e., that $T_{\mu\;\:;\nu}^{\;\:\nu}=0$, and calculate $\Phi$ from the contracted Bianchi identities, arriving at the Einstein equations. However, one cannot derive condition $T_{\mu\;\:;\nu}^{\;\:\nu}=0$ from thermodynamics of spacetime. Thus, it represents an additional, nontrivial assumption. A simpler way to specify $\Phi$, without introducing any new conditions, is by taking a trace of the equations. Then, we obtain
\begin{equation}
R_{\mu\nu}-\frac{1}{4}Rg_{\mu\nu}=\frac{2\pi k_B}{\eta\hbar c}\left(T_{\mu\nu}-\frac{1}{4}Tg_{\mu\nu}\right).
\end{equation}
Lastly, we define the Newton's gravitational constant in terms of $\eta$ as $G=k_Bc^3/4\hbar\eta$ (by requiring the correct Newtonian limit of the equations). This implies that the entropy of Rindler horizon corresponds to Bekenstein entropy
\begin{equation}
S_{\text{horizon}}=\eta\mathcal{A}=\frac{k_Bc^3\mathcal{A}}{4G\hbar}=S_{Bekenstein}.
\end{equation}
Hence, the recovery of gravitational dynamics from thermodynamics requires that black hole horizons and Rindler horizons (and, indeed, local causal horizons of any construction) have the same entropy per unit area, $\eta=k_Bc^3/4G\hbar$. Conversely, one might take the view that traceless Einstein equations together with the first law of thermodynamics imply universal entropy density for any causal horizon (for a detailed discussion, see~\cite{Jacobson:2003,Jacobson:2019}).

In total, we have obtained the following equations for gravitational dynamics
\begin{equation}
R_{\mu\nu}-\frac{1}{4}Rg_{\mu\nu}=\frac{8\pi G}{c^4}\left(T_{\mu\nu}-\frac{1}{4}Tg_{\mu\nu}\right),
\end{equation}
which are identical to EoMs of UG. Furthermore, the behavior of the Rindler wedge set-up under Weyl transformations seems to suggest that these are in fact EoMs of WTG in the unimodular gauge~\cite{Hammad:2019} (we will address this observation in a future work, the effect of conformal transformations in thermodynamics of spacetime is also discussed in~\cite{Tiwari:2006}). In any case, the cosmological constant appears as an arbitrary integration constant (whether we assume local energy-momentum conservation or not) and its value can vary between solutions, exactly as in unimodular theories.

\subsection{Thermodynamics of local causal diamonds}

While seminal papers concerning thermodynamics of spacetime considered causal horizons realized as local Rindler wedges~\cite{Jacobson:1995,Padmanabhan:2010,Chirco:2010}, there are some drawbacks to this approach. First, the boundary of the 2-dimensional spacelike patch $\mathcal{B}$ used to define the local Rindler wedge is chosen arbitrarily. Second, the finite strip of a Rindler horizon we considered does not form a boundary of some interior region causally disconnected from the exterior. Therefore, it is not quite clear whether it should posses entanglement entropy~\cite{Svesko:2018} (assuming one chooses to interpret Bekenstein entropy in terms of entanglement). Moreover, it has been argued that Clausius entropy flux associated with a local Rindler wedge cannot be correctly interpreted in terms of quantum von Neumann entropy~\cite{Carroll:2016}. Lastly, the entropy of Rindler horizons, in contrast with black holes, does not acquire quantum corrections logarithmic in area~\cite{Solodukhin:2011}. Consequently, thermodynamics of local Rindler horizons cannot provide any insights into low energy quantum gravity effects~\cite{Alonso:2020b,Solodukhin:2011}. Fortunately, all the above mentioned problems disappear when one replaces local Rindler horizons with spherical local causal horizons (in principle, closed horizons of any shape would be suitable, but calculations quickly become unmanageable beyond spherical symmetry). Then the boundary of the horizon's spatial cross-section is an approximate 2-sphere determined by a single length scale. The 2-sphere encloses an interior region that can plausibly posses entropy due to its entanglement with the exterior and von Neumann and Clausius entropy of the matter inside turn out to be in agreement~\cite{Alonso:2020a}.

\subsubsection{Geodesic local causal diamonds}

\begin{figure}[tbp]
\centering
\includegraphics[width=.45\textwidth,origin=c,trim={0.1cm 2.4cm 36.7cm 1.5cm},clip]{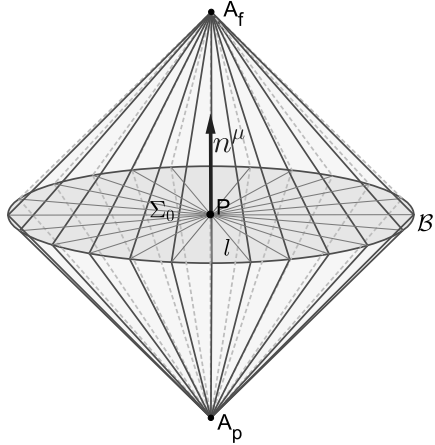}
\caption{\label{diamond} A schematic picture of a GLCD with angular coordinate $\theta$ suppressed. Diamond's base $\Sigma_0$ is a spacelike geodesic ball, formed by geodesics of parameter length $l$ sent out from point $P$ (represented by the grey lines inside the base). Boundary $\mathcal{B}$ of $\Sigma_0$ is approximately a two-sphere. The ball is orthogonal to a timelike vector $n^\mu$. The tilted lines represent geodesic generators of the diamond's null boundary. The generators all start from past apex $A_p$ (corresponding to coordinate time $-l/c$) and again converge together in future apex $A_f$ (coordinate time $l/c$). Thus, the diamond's base is the spatial cross section of the future domain of dependence of $A_p$ at coordinate time $t=0$ and, likewise, the cross section of the past domain of dependence of $A_f$.}
\end{figure}

In the following, we will consider a local spherical horizon constructed as small geodesic local causal diamonds (GLCD)\footnote{Note that one can choose to work with null cones instead. However, since the results are equivalent in both cases~\cite{Svesko:2018,Svesko:2019} (at least in the semiclassical setting), we concentrate on thermodynamics of GLCD's in this work.} To construct a GLCD centered at point $P$, consider an arbitrary unit timelike vector $n^{\mu}$ and send out geodesics of parameter length $l$ in all directions orthogonal to $n^{\mu}$. These geodesics form a spacelike geodesic ball $\Sigma_0$. The GLCD is then defined as the spacetime region causally determined by $\Sigma_0$ (see Fig.~\ref{diamond}). There exists an approximate (up to $O\left(l^3\right)$ terms) spherically symmetric conformal isometry preserving the GLCD generated by a conformal Killing vector~\cite{Jacobson:2015}
\begin{equation}
\zeta=\frac{1}{2l}\left(\left(l^2-t^2-r^2\right)\frac{\partial}{\partial t}-2rt\frac{\partial}{\partial r}\right).
\end{equation}
It is easy to see that the null boundary of the GLCD is a conformal Killing horizon.

\subsubsection{Physical process derivation}

We discuss two different derivations of gravitational dynamics from thermodynamics of GLCD's, one developed by Svesko~\cite{Svesko:2019} and the other by Jacobson~\cite{Jacobson:2015}. We present both approaches in a slightly modified form to emphasize their unimodular features. The first method is based on studying a physical process, namely the flux of Clausius entropy entering the past GLCD's horizon~\cite{Svesko:2019}. We can calculate it in the same way as for a local Rindler wedge~\cite{Baccetti:2013,Alonso:2020a}. The total entropy flux between times $t<0$ and $0$ (corresponding to the bifurcation surface) equals~\cite{Baccetti:2013}
\begin{equation}
\Delta S_{\text{Clausius}}=-\frac{2\pi k_B}{\hbar c}\int_{-t}^{0}\text{d}t' t'\int_{\mathcal{S}\text{d}^2\mathcal{A}\left(t\right)} T_{\mu\nu}\xi^{\mu}\xi^{\nu},
\end{equation}
where $\mathcal{B}\left(t\right)$ is a spatial cross-section of the GLCD's null boundary at time $t$ (an approximate $2$-sphere of radius $l-ct$) and $\xi^{\mu}$ denotes a future-oriented null vector field orthogonal to $\Sigma$. We could now again assume that the horizon possesses entropy proportional to its area. However, the GLCD settings allow us to treat a more general entropy formula~\cite{Svesko:2018,Svesko:2019}\footnote{Thermodynamics of local Rindler wedges has also been extended in this way~\cite{Padmanabhan:2010,Jacobson:2012}. However, all the proposed extensions suffer drawbacks which are not encountered for GLCD's and null cones~\cite{Svesko:2018}. These drawbacks can be traced back to the lack of spherical symmetry for Rindler wedges and absence of a well defined ``interior region''.}
\begin{equation}
S_{\text{horizon}}\left(t\right)=\int_{\mathcal{B}\left(t\right)}S_{\mu\nu}\epsilon^{\mu\nu}\text{d}^2\mathcal{A},
\end{equation}
where $\epsilon^{\mu\nu}=m^{\mu}n^{\nu}-n^{\mu}m^{\nu}$ denotes a bi-normal to $\mathcal{B}\left(t\right)$ (with $m=\partial/\partial r$ being a unit radial vector) and $S_{\mu\nu}$ is some entropy density tensor. This represents a generalization of Bekenstein entropy which we recover by choosing~\cite{Wald:1994}
\begin{equation}
S_{\mu\nu}=\frac{k_Bc^3}{16\hbar G}\left(g_{\mu\rho}g_{\nu\sigma}-g_{\mu\sigma}g_{\nu\rho}\right)\zeta^{\rho;\sigma}.
\end{equation}
More generally, one might consider entropy density defined by the Wald's Noether charge prescription~\cite{Wald:1994}. For any theory whose Lagrangian density $\mathcal{L}$ does not depend on derivatives of the Riemann tensor, it holds~\cite{Svesko:2018,Wald:1994}
\begin{equation}
S_{\mu\nu}=\frac{k_Bc^3}{8\hbar G}\left(-2P_{\mu\nu\rho\sigma}\zeta^{\rho;\sigma}+4P_{\mu\nu\:\:\sigma;\rho}^{\;\;\;\,\rho}\zeta^{\sigma}\right),
\end{equation}
where $P^{\mu\nu\rho\sigma}=\left(16\pi G/c^4\right)\partial\mathcal{L}/\partial R_{\mu\nu\rho\sigma}$ depends on the metric and the Riemann tensor. For the Wald entropy difference between times $t=-\varepsilon$ such that $0<\varepsilon\ll l/c$ and $t=0$ we obtain~\cite{Svesko:2019}
\begin{align}
\nonumber \Delta S_{\text{horizon}}\left(\varepsilon\right)=&\int_{\mathcal{B}\left(0\right)}S_{\mu\nu}\zeta^{\nu;\mu}\text{d}^2\mathcal{A}-\int_{\mathcal{B}\left(-\varepsilon\right)}S_{\mu\nu}\zeta^{\nu;\mu}\text{d}^2\mathcal{A}\approx-\frac{k_Bc^3}{4\hbar G}\int_{-\varepsilon}^{0}\text{d}t\int_{\mathcal{B}\left(t\right)}\text{d}^2\mathcal{A}\xi^{\mu}\\
&\left[-2P^{\mu\nu\rho\sigma}_{\quad\;\;\;;\nu\rho}\zeta_{\sigma}-\left(P_{\mu\nu\rho\sigma}^{\quad\;\;\;;\rho}+P_{\mu\sigma\rho\rho}^{\quad\;\;\;;\rho}\right)\zeta^{\nu;\sigma}+P_{\mu\nu\rho\sigma}\left(R^{\sigma\rho\nu\iota}\zeta_{\iota}+f^{\nu\rho\sigma}\right)\right],
\end{align}
where term $R_{\rho\mu\sigma\nu}\zeta^{\sigma}$ comes from the Killing identity and $f_{\rho\mu\nu}$ accounts for the fact that this identity is not satisfied by $\zeta^{\mu}$. This happens, on one side, because $\zeta^{\mu}$ is only a conformal Killing vector even in flat spacetime. On the other side, $f_{\rho\mu\nu}$ includes terms appearing due to the isometry generated by $\zeta^{\mu}$ being only approximate in curved spacetime (up to $O\left(l^3\right)$ curvature dependent terms). For the same reasons, the term $-\left(P_{\mu\nu\rho\sigma}^{\quad\;\;\;;\rho}+P_{\mu\sigma\rho\rho}^{\quad\;\;\;;\rho}\right)\zeta^{\nu;\sigma}$ does not vanish, although it would be the case for a true Killing vector. However, it has been shown that the spherical symmetry combined with correcting the definition of $\zeta^{\mu}$ by higher order terms in $l$ leads to~\cite{Svesko:2019}
\begin{equation}
-\frac{k_Bc^3}{4\hbar G}\int_{-\varepsilon}^{0}\text{d}t\int_{\mathcal{B}\left(t\right)}\text{d}^2\mathcal{A}\xi^{\mu}\left[-\left(P_{\mu\nu\rho\sigma}^{\quad\;\;\;;\rho}+P_{\mu\sigma\rho\rho}^{\quad\;\;\;;\rho}\right)\zeta^{\nu;\sigma}+P_{\mu\nu\rho\sigma}f^{\nu\rho\sigma}\right]=\Delta S_{\text{flat}},
\end{equation} 
corresponding to the change in Wald entropy of the GLCD in a flat spacetime. This contribution can be understood as irreversible change of entropy~\cite{Svesko:2018,Svesko:2019}. Since the heat flux compensates only the reversible part of the entropy flux~\cite{Chirco:2010,Svesko:2018}, the entropy balance
\begin{equation}
\Delta S_{\text{Wald}}-\Delta S_{\text{flat}}+\Delta S_{\text{Clausius}}=0,
\end{equation}
requires
\begin{equation}
\int_{-\varepsilon}^{0}\text{d}t\int_{\mathcal{B}\left(t\right)}\text{d}^2\mathcal{A}\xi^{\mu}\left(\frac{k_Bc^3}{2\hbar G}P_{\mu\nu\rho\sigma}^{\quad\;\;\;;\nu\rho}\zeta^{\sigma}-\frac{k_Bc^3}{4\hbar G}P_{\mu\nu\rho\sigma}R^{\sigma\rho\nu\iota}\zeta_{\iota}-t\frac{2\pi k_B}{\hbar c}T_{\mu\nu}\xi^{\mu}\xi^{\nu}\right)=0.
\end{equation}
Approximating $P^{\mu\nu\rho\sigma}$, $R_{\sigma\rho\nu\iota}$ and $T_{\mu\nu}$ by their values in the GLCD's origin $P$ allows us to carry out the integration\footnote{In the original paper~\cite{Svesko:2019}, the integration is not considered. However, we choose to perform it as it directly leads to traceless equations of motion without the presence of any undetermined terms.}
\begin{equation}
2\pi l^2\varepsilon^2\frac{k_Bc^3}{4\hbar G}\left(2P_{\mu\sigma\rho\nu}^{\quad\;\;\;;\sigma\rho}-P_{\mu\sigma\rho\iota}R^{\iota\rho\sigma}_{\quad\nu}+\frac{8\pi G}{c^4}T_{\mu\nu}\right)\left(n^{\mu}n^{\nu}+\frac{1}{3}\left(g^{\mu\nu}+n^{\mu}n^{\nu}\right)\right)+O\left(\varepsilon^3\right)=0.
\end{equation}
Since a GLCD can be constructed with respect to any timelike direction, the previous equation must hold for every unit timelike vector $n^{\mu}$ defined in point $P$. The entropy balance for every such GLCD centered in $P$ thus demands
\begin{equation}
P_{(\mu\vert\sigma\rho\iota}R^{\iota\rho\sigma}_{\quad\vert\nu)}-2P_{(\mu\vert\sigma\rho\vert\nu)}^{\quad\quad\;\;\;;\sigma\rho}-\frac{1}{4}g^{\kappa\lambda}\left(P_{\kappa\sigma\rho\iota}R^{\iota\rho\sigma}_{\quad\lambda}-2P_{\kappa\sigma\rho\lambda}^{\quad\;\;\;;\sigma\rho}\right)g_{\mu\nu}=\frac{8\pi G}{c^4}\left(T_{\mu\nu}-\frac{1}{4}Tg_{\mu\nu}\right).
\end{equation}
The equivalence principle ensures that this equation can be derived in every spacetime point. In the end, we have recovered traceless equations of motion for any local theory of gravity whose Lagrangian is independent of the derivatives of the Riemann tensor and satisfies $P^{\mu\nu\rho\sigma}=\left(16\pi G/c^4\right)\partial\mathcal{L}/\partial R_{\mu\nu\rho\sigma}$. However, although our starting point was the entropy density derived from a fully diffeomorphism invariant Lagrangian with a constant term corresponding to the cosmological constant, only the traceless part of equations of motion has been recovered. Even if we impose local energy-momentum conservation to recover the full equations of motion, the cosmological term only appears as an integration constant unrelated to the parameter in the original Lagrangian. This behavior points towards a unimodular theory of gravity rather than a fully diffeomorphism invariant one. Thus, the unimodular character of thermodynamically derived equations of gravitational dynamics persists even when one starts from a more general Wald expression for entropy density (this remains true even when quantum gravity corrections to entropy are taken into account~\cite{Alonso:2020b}). The role of Weyl symmetry in this case will be discussed in a future work.

\subsubsection{Maximal vacuum entanglement hypothesis derivation}

The last derivation we consider is based on studying a small variation of the GLCD set-up rather than on the entropy flux. Such a variation is expected to behave in accord with the maximal vacuum entanglement hypothesis: ``When the geometry and quantum fields are simultaneously varied from maximal symmetry, the entanglement entropy in a small geodesic ball is maximal at fixed volume~\cite{Jacobson:2015}.'' This approach has the advantage of describing both the geometry and matter entropy in terms of quantum entanglement, whereas the previously discussed derivations treats matter entropy as classical Clausius entropy (although Clausius and entanglement descriptions of matter entropy are in fact equivalent~\cite{Alonso:2020a}).

Consider a small GLCD in a maximally symmetric spacetime in vacuum. The maximal vacuum entanglement hypothesis states that the entanglement entropy of the GLCD is then maximal~\cite{Jacobson:2015}. Consequently, a variation of entanglement entropy vanishes to first order. Under a simultaneous variation of spacetime geometry and matter fields, the change in entanglement entropy has two components. The change due to variation of the geometry is, as discussed in subsection~\ref{Jacobson} universal and proportional to the area, $\delta S_{\text{horizon}}=\eta\delta\mathcal{A}$. The variation of the area of 2-sphere $\mathcal{B}$ at fixed volume of geodesic ball $\Sigma_0$\footnote{For a physical justification of this condition, see~\cite{Bueno:2017,Jacobson:2019b,Svesko:2019}.} reads
\begin{equation}
\delta\mathcal{A}=-\frac{4\pi}{15}l^4\left(G_{\mu\nu}n^{\mu}n^{\nu}-\Lambda+\delta\Lambda\right),
\end{equation}
where $\Lambda$ is the cosmological constant of the original maximally symmetric spacetime and $\delta\Lambda$ its variation inside the GLCD (in general dependent on the position in spacetime and the length scale $l$). Varying cosmological constant has already been introduced in the context of thermodynamics of GLCD's in GR~\cite{Jacobson:2019,Jacobson:2019b}, but it becomes especially natural in the light of its relation to UG\footnote{Of course, one could carry out the derivation without assuming varying cosmological constant and return to the issue once the unimodular nature of the dynamics is shown (we have followed this approach in an earlier paper~\cite{Alonso:2020a}). However, for the sake of clarity, we take advantage of the insights we gained by the previously discussed methods.}. In UG (and WTG) two solutions of EoMs can be characterised by different $\Lambda$, as it is merely an arbitrary integration constant.

The second component of the change in the entanglement entropy appears due to variation of matter fields away from vacuum. The vacuum state inside the GLCD can be expressed as a thermal density matrix, $\rho_{\text{vac}}=\exp\left(-K/T_U\right)$, where $K$ is known as the modular Hamiltonian and $T_U=\hbar c/2\pi k_B$ is the Unruh temperature~\cite{Jacobson:2019,Jacobson:2015}. Knowing the behavior of $K$ under variation of the matter fields then allows us to calculate the corresponding change in von Neumann entropy. For conformal matter fields, $\delta K$ turns out to be local and proportional to variation of the energy-momentum tensor expectation value, $\delta\langle T_{\mu\nu}\rangle$. This is no longer true for non-conformal fields. Nevertheless, if they posses a fixed UV point, the variation of $K$ is a sum of the term proportional to $\delta\langle T_{\mu\nu}\rangle$ and an additional scalar function~\cite{Speranza:2016,Casini:2016}. In total, variation of the von Neumann entropy of matter fields obeys
\begin{equation}
\delta S_{\text{matter}}=\frac{2\pi k_B}{\hbar c}\frac{4\pi l^4}{15}\left(\delta\langle T_{\mu\nu}\rangle n^{\mu}n^{\nu}+\delta X\right),
\end{equation}
where $X$ is a spacetime scalar that in general depends on $l$. For conformal fields it holds $X=0$.

In total, the entanglement equilibrium condition $\delta S_{\text{horizon}}+\delta S_{\text{matter}}=0$ states
\begin{equation}
-\frac{4\pi}{15}l^4\eta\left(G_{\mu\nu}n^{\mu}n^{\nu}-\Lambda+\delta\Lambda\right)+\frac{2\pi k_B}{\hbar c}\frac{4\pi l^4}{15}\left(\delta\langle T_{\mu\nu}\rangle n^{\mu}n^{\nu}+\delta X\right)=0.
\end{equation}
Since the unit timelike vector $n^{\mu}$ is arbitrary, it must hold
\begin{equation}
G_{\mu\nu}+\Lambda g_{\mu\nu}-\delta\Lambda g_{\mu\nu}-\frac{2\pi k_B}{\hbar c\eta}\left(\delta\langle T_{\mu\nu}\rangle-\delta X g_{\mu\nu}\right)=0.
\end{equation}
Taking a trace of the equations, we find a condition on the variation of the cosmological constant
\begin{equation}
\delta\Lambda=\frac{1}{4}R-\Lambda+\frac{2\pi k_B}{\hbar c\eta}\left(\frac{1}{4}\delta\langle T\rangle-\delta X\right),
\end{equation}
and, setting $G=c^3/4k_B\hbar$ as before, we obtain the traceless equations equivalent to EoMs of UG
\begin{equation}
R_{\mu\nu}-\frac{1}{4}R g_{\mu\nu}=\frac{8\pi G}{c^4}\left(\delta\langle T_{\mu\nu}\rangle-\frac{1}{4}\delta\langle T\rangle g_{\mu\nu}\right).
\end{equation}
Once again, the strong equivalence principle ensures that the equations are valid throughout the spacetime. The main new feature of this approach is that on the right hand side appears an expectation value of energy-momentum tensor of quantum fields. Hence, the equations we obtained should be interpreted as semiclassical EoMs of UG.

To conclude, we remark that if all the fields present in the spacetime are conformal, then $\delta\langle T\rangle=\delta X=0$, $R=4\Lambda$ and, consequently, $\delta\Lambda=0$. In this way, we see that a variation of the cosmological constant occurs if and only if one deals with a non-conformal field theory. Interestingly, one of the explanations of the origin of $\Lambda$ proposed in the context of UG relies on the effects of non-conformal fields in the early universe~\cite{Perez:2018}. However, we presently do not know whether this or some similar proposal can be connected with the maximal vacuum entanglement hypothesis.\footnote{Let us note that an interpretation of $\delta X$ within the unimodular framework unrelated to the cosmological constant has been recently put forward~\cite{Tiwari:2021}. This proposal relates $\delta X$ with the matter Lagrangian ambiguity $l_{\text{M}}$ (see Eq.~\eqref{action 2} in section~\ref{UG} and the following discussion).}

\section{Discussion}
\label{discussion}

All three ways to derive gravitational EoMs from thermodynamics we reviewed in this work lead to unimodular theories of gravity (this is in fact true for any thermodynamic approach known to the authors). In each case we found traceless EoMs that only constrain divergence of the energy-momentum tensor to be equal to a gradient of some scalar. If we impose divergence-free energy-momentum tensor as an additional condition, we recover equations of the same form as EoMs of the corresponding fully diffeomorphism invariant theory. However, the cosmological constant appears as an integration constant of arbitrary value, in correspondence with its behavior in unimodular theories of gravity. Notably, even when we use thermodynamics of spacetime to study semiclassical gravity or modified gravitational dynamics, the result possesses an unimodular character. As we explored in a previous work~\cite{Alonso:2020b}, this remains true even when we consider low energy quantum gravity effects. In this case it even appears that the equivalence between unimodular and fully diffeomorphism invariant gravitational dynamics breaks down.

Moreover, based on the results available in the literature, thermodynamically derived gravitational dynamics seems to be invariant under Weyl transformations. This suggests that thermodynamic derivations lead directly to WTG rather than UG (or to appropriate generalizations of these theories). However, we leave a systematic study of this possibility for a future work.
 
To conclude, thermodynamics of spacetime, when considered as a way to gain insight into the nature of gravitational dynamics, clearly point to its unimodular (or rather Weyl transverse) character. Interestingly, UG, and especially WTG, offer some advantages over GR in the behaviour of the cosmological constant, while reproducing all of its classical predictions. It would be interesting to check whether thermodynamic methods have something to say about the origin of the cosmological constant. Likewise, the suggested breakdown of the equivalence between unimodular and fully diffeomorphism invariant gravitational dynamics due to quantum gravity effects deserves further attention.

\section*{Acknowledgments}

AA-S is supported by the ERC Advanced Grant No. 740209. M.L. is supported by the Charles University Grant Agency project No. GAUK 297721. This work is also partially supported by Project. No. MICINN PID2020-118159GB-C44.

\end{document}